\documentclass[preprint]{imsart}

\usepackage{amsthm,amsmath,amsfonts}
\RequirePackage[colorlinks,citecolor=blue,urlcolor=blue]{hyperref}
\usepackage[round]{natbib}
\usepackage[english]{babel}
\usepackage{latexsym}
\usepackage{subfigure}
\usepackage{epsfig}
\usepackage{color,soul}
\usepackage{moreverb}
\usepackage{mathtools}
\usepackage{tikz, color}
\usepackage{enumerate,linegoal}
\usepackage{calc}
\usepackage{latexsym}
\usepackage{amsmath}
\usepackage{amsfonts}
\usepackage{amssymb}
\usepackage{bm}
\usepackage{psfrag}
\usepackage{graphicx}
\usepackage{epstopdf}
\usepackage{framed}
\DeclareMathOperator*{\argmin}{arg\,min}

\usetikzlibrary{matrix}

\startlocaldefs
\newcommand{\expect}{\operatorname{E}}

\newcommand{\rd}{\mathrm{d}}
\newcommand{\rds}{\,\rd}
\renewcommand{\eqref}[1]{(\ref{#1})}

\newcommand{\tp}{^{\top}}

\endlocaldefs

\begin{document}

\begin{frontmatter}

\title{Nonlinear quantile mixed models}
\runtitle{Nonlinear quantile mixed models}

\begin{aug}
\author{\fnms{Marco} \snm{Geraci}\corref{}\thanksref{t1}\ead[label=e1]{geraci@mailbox.sc.edu}}

\thankstext{t1}{Corresponding author: Marco Geraci, Department of Epidemiology and Biostatistics, Arnold School of Public Health, University of South Carolina, 915 Greene Street, Columbia SC 29209, USA. \printead{e1}}

\runauthor{M. Geraci}

\affiliation{University of South Carolina\thanksmark{t1}}

\end{aug}

\begin{abstract}
\quad In regression applications, the presence of nonlinearity and correlation among observations offer computational challenges not only in traditional settings such as least squares regression, but also (and especially) when the objective function is non-smooth as in the case of quantile regression. In this paper, we develop methods for the modeling and estimation of nonlinear conditional quantile functions when data are clustered within two-level nested designs. This work represents an extension of the linear quantile mixed models of Geraci and Bottai (2014, \textit{Statistics and Computing}). We develop a novel algorithm which is a blend of a smoothing algorithm for quantile regression and a second order Laplacian approximation for nonlinear mixed models. To assess the proposed methods, we present a simulation study and two applications, one in pharmacokinetics and one related to growth curve modeling in agriculture.
\end{abstract}



\end{frontmatter}

\section{Introduction}
\label{sec:1}
Quantile regression analysis of clustered data is a very active area of research. Since the seminal work of \cite{koenker1978} on methods for cross-sectional observations, there have been a number of proposals on how to accommodate for the dependency induced by clustered (e.g., longitudinal) designs. As outlined by \cite{geraci2014} and then extensively reviewed by \cite{marino2015}, approaches to linear quantile regression with clustered data can be classified into distribution-free and likelihood-based approaches. The former include fixed effects \citep{koenker2004, lamarche2010, galvao2010, galvao2011} and weighted \citep{lipsitz1997, fu2012} approaches. The latter are mainly based on the asymmetric Laplace (AL) density \citep{geraci2007, geraci2014, yuan2010, farcomeni2012} or other, usually flexible, parametric distributions \citep[for example, ][]{rigby2005,reich2010,noufaily2013}. A different classification can be made into approaches that include cluster-specific effects \citep[e.g.,][]{koenker2004,geraci2014} and those that ignore or remove them \citep{lipsitz1997,canay2011,parente2016}.

In some applications, the assumption of linearity may not be appropriate. This is often the case in, for example, pharmacokinetics \citep{lindsey} and growth curve modeling \citep{panik}. Contributions to statistical methods for nonlinear mean regression when data are clustered can be found in the literature of mixed-effects modeling \citep{lindstrom_bates,pinheiro1995,pinheiro_bates} as well as generalized estimating equations \citep{davidian_1995,davidian_2003,contreras,vonesh_etal}. We examined the statistical literature on \emph{parametric} nonlinear quantile regression functions with clustered data (thus, in our review, we did not consider smoothing for nonparametric quantile functions). To the best of our knowledge, there seem to be only a handful of published articles. \cite{karlsson2008} considered nonlinear longitudinal data and proposed weighting the standard quantile regression estimator \citep{koenker1978} with pre-specified weights. \cite{wang2012}, taking her cue from \cite{geraci2007}, used the AL distribution to define the likelihood of a Bayesian nonlinear quantile regression model. \cite{huang2016} proposed a Bayesian joint model for time-to-event and longitudinal data. An approach based on copulas is developed by \cite{chen_etal}. Finally, \cite{oberhofer2016} established the consistency of the $L_{1}$-norm nonlinear quantile estimator under weak dependency. None of the cited papers provides an approach to modeling and estimation of nonlinear quantile functions with random effects in a frequentist framework. Such an approach is desirable when the correlation between measurements is modeled by means of random effects, but the parameters of interest are assumed to be fixed.

In this paper, we propose an extension of \citeauthor{geraci2014}'s (\citeyear{geraci2014}) linear quantile mixed model (LQMM) to the nonlinear case. In Section 2, we briefly outline the LQMM approach and, in Section 3, we introduce the nonlinear quantile mixed-effects model or nonlinear quantile mixed model (NLQMM) for short. Estimation is carried out using a novel algorithm which is a combination of a smoothing algorithm for quantile regression and a second order Laplacian approximation for nonlinear mixed models. We then carry out a simulation study in Section 4 to assess the performance of the proposed methods. In Section 5, we consider an application of NLQMM to pharmacokinetics and growth curves modeling. We conclude with some remarks in Section 6.

\section{Linear quantile mixed models}
\label{sec:2}
Consider data from a two-level nested design in the form $(\mathbf{x}_{ij}\tp,\mathbf{z}_{ij}\tp,y_{ij})$, for $j=1,\ldots , n_{i}$ and $i=1,\ldots , M$, $N = \sum_i n_i$, where $\mathbf{x}_{ij}\tp$ is the $j$th row of a known $n_{i}\times p$ matrix $\mathbf{X}_i$, $\mathbf{z}_{ij}\tp$ is the $j$th row of a known $n_{i}\times q$ matrix $\mathbf{Z}_i$ and $y_{ij}$ is the $j$th observation of the response vector $\mathbf{y}_i = (y_{i1},\ldots,y_{in_{i}})\tp$ for the $i$th cluster. Throughout the paper, the covariates $x$ and $z$ are assumed to be given. The $n \times 1$ vector of ones will be denoted by $\mathbf{1}_{n}$, the $n \times n$ identity matrix by $\mathbf{I}_{n}$, and the $m \times n$ matrix of zeros by $\mathbf{O}_{m \times n}$.

In a distribution-free approach, the linear quantile regression model for clustered (or panel) data \citep[e.g.,][]{koenker2004,abrevaya2008,bache2013} can be specified as
\begin{equation}\label{eq:1}
Q_{y_{ij}}(\tau) = \mathbf{x}_{ij}\tp\bm\beta_{\tau} + \mathbf{z}_{ij}\tp\bm\delta_{\tau,i},
\end{equation}
where $0 < \tau < 1$ is the given quantile level, $\bm\beta_{\tau}$ is a $p \times 1$ vector of coefficients common to all clusters, while the $q \times 1$ vector $\bm\delta_{\tau,i}$ may vary with cluster. All the parameters are $\tau$-specific, although the cluster-specific effects are often specified simply as pure location-shift effects \citep{koenker2004,lamarche2010}. Fitting can be achieved by solving the classical $L_{1}$-norm regression problem
\begin{equation}\label{eq:2}
\min_{\bm\beta,\bm\delta} \sum_{i=1}^{M}\sum_{j=1}^{n_{i}} \rho_\tau\left(y_{ij} - \mathbf{x}_{ij}\tp\bm\beta - \mathbf{z}_{ij}\tp\bm\delta_{i}\right) + \sum_{i=1}^{M}\mathcal{P}(\bm\delta_{i}),
\end{equation}
where $\rho_\tau(r)=r\left\{\tau-I(r < 0)\right\}$ is the loss function and $I$ denotes the indicator function. The penalty $\mathcal{P}$ on the cluster-specific effects controls the variability introduced by a large number of parameters $\bm\delta_{i}$ and is usually based on the $L_{1}$-norm \citep{koenker2004, lamarche2010, bache2013}.

To mimic the minimization problem (\ref{eq:2}) in a likelihood framework, \cite{geraci2014} introduced the convenient assumption that the responses $y_{ij}$, $j=1,\ldots , n_{i}$, $i=1,\ldots,M$, conditionally on a $q\times1$ vector of random effects $\mathbf{u}_i$, independently follow the asymmetric Laplace (AL) density
\[
p(y_{ij}|\mathbf{u}_{i}) = \frac{\tau(1-\tau)}{\sigma_{\tau}}\exp\left\{-\frac{1}{\sigma_{\tau}}\rho_\tau\left(y_{ij}-\mu_{\tau,ij}\right)\right\},
\]
with location and scale parameters given by $\mu_{\tau,ij} = \mathbf{x}_{ij}\tp\bm\beta_{\tau} + \mathbf{z}_{ij}\tp\mathbf{u}_{i}$ and $\sigma_{\tau}$, respectively, which we write as $y_{ij} \sim \mathcal{AL}\left(\mu_{\tau,ij}, \sigma_{\tau}\right)$. (The third parameter of the AL is the skew parameter $\tau \in (0,1)$ which, in this model, is fixed and defines the quantile level of interest.) Also, they assumed that $\mathbf{u}_{i} = \left(u_{i1},\ldots,u_{iq}\right)\tp$, for $i=1,\ldots,M$, is a random vector independent from the model's error term with mean zero and $q\times q$ variance-covariance matrix $\bm\Sigma_{\tau}$. Note that all the parameters are $\tau$-dependent. The random effects vector $\mathbf{u}$ depends on $\tau$ through the variance-covariance matrix. If we let $\mathbf{u} = (\mathbf{u}_{1}\tp,\ldots,\mathbf{u}_{M}\tp)\tp$ and $\mathbf{y} = (\mathbf{y}_{1}\tp,\ldots,\mathbf{y}_{M}\tp)\tp$, the joint density $p\left(\mathbf{y},\mathbf{u}\right) \equiv p\left(\mathbf{y}|\mathbf{u}\right)p\left(\mathbf{u}\right)$ based on $M$ clusters for the $\tau$th linear quantile mixed model (LQMM) is given by \citep{geraci2014}
\begin{equation*}
p\left(\mathbf{y},\mathbf{u}\right) = \left\{\frac{\tau(1-\tau)}{\sigma_{\tau}}\right\}^{N}
\prod_{i=1}^{M} \exp\left\{-\frac{1}{\sigma_{\tau}}\sum_{j=1}^{n_{i}} \rho_\tau\left(y_{ij}-\mu_{\tau,ij}\right)\right\}p\left(\mathbf{u}_{i}\right).
\end{equation*}

\cite{geraci2014} proposed estimating LQMMs through a combination of Gaussian quadrature and non-smooth optimization. They approximated the marginal (over the random effects) log-likelihood using the rule
\begin{equation}\label{eq:3}
\ell_{\mathrm{GQ}}(\bm\beta_{\tau},\bm\Sigma_{\tau},\sigma_{\tau}|\mathbf{y})= \sum_{i}^{M}\log\left\{\sum_{k_1=1}^{K}\cdots\sum_{k_q=1}^{K}p\left(\mathbf{y}_{i}| \mathbf{v}_{k_1,\ldots,k_q}\right) \prod_{l=1}^{q}w_{k_{l}}\right\},
\end{equation}
with $\mathbf{v}_{k_1,\ldots,k_q}=(v_{k_1},\ldots,v_{k_q})\tp$, where $v_{k_l}$ and $w_{k_l}$, $k_l = 1,\ldots,K$, $l=1,\ldots,q$, denote, respectively, the abscissas and weights of the (one-dimensional) Gaussian quadrature. In principle, one can consider different distributions for the random effects, which may be naturally linked to different quadrature rules (or penalties). For example, it is immediate to verify that the double exponential distribution confers robustness to the model and is akin to a Gauss-Laguerre quadrature \citep{geraci2014}. Alternatively, one can avoid parametric assumptions altogether \citep{alfo2016}. Throughout this paper we assume that the random effects are normally distributed and we do not explore this issue any further. However, this assumption can be relaxed and the developments that follow can be modified as appropriate.

\section{Nonlinear quantile mixed models}
\label{sec:3}

\subsection{The model}
\label{sec:3.1}
We consider the nonlinear quantile regression function
\begin{equation}\label{eq:4}
Q_{y_{ij}|\mathbf{u}_{i}}(\tau) = f\left(\bm\phi_{\tau,ij},\mathbf{x}_{ij}\right),
\end{equation}
where $f$ is a nonlinear, smooth function of the $s \times 1$ random parameter $\bm\phi_{\tau,ij} = \mathbf{F}_{ij}\bm\beta_{\tau} + \mathbf{G}_{ij}\mathbf{u}_{i}$, $\mathbf{F}_{ij}$ and $\mathbf{G}_{ij}$ are two given design matrices of dimensions $s \times p$ and $s \times q$, respectively, which in general contain elements of the covariates $\mathbf{x}_{ij}$.

To stress the functional dependence of the quantiles on the $p \times 1$ fixed parameter $\bm\beta_{\tau}$ and on the $q \times 1$ random parameter $\mathbf{u}_{i}$, we write $f\left(\bm\phi_{\tau,ij},\mathbf{x}_{ij}\right) \equiv f_{ij}\left(\bm\beta_{\tau},\mathbf{u}_{i}\right)$. For estimation purposes, model~\eqref{eq:4} can be equivalently written as
\begin{equation}\label{eq:5}
y_{ij} = f_{ij}\left(\bm\beta_{\tau},\mathbf{u}_{i}\right) + \epsilon_{\tau,ij},
\end{equation}
conditionally on $\mathbf{u}_{i}$, where $\epsilon_{\tau,ij} \sim \mathcal{AL}\left(0, \sigma_{\tau}\right)$. Moreover, we assume $\mathbf{u}_{i} \sim \mathcal{N}\left(\mathbf{0}, \bm\Sigma_{\tau}\right)$, independently from $\epsilon_{ij}$.

Note the similarities and dissimilarities between the proposed model \eqref{eq:5} and the traditional nonlinear mixed-effects (NLME) model
\begin{equation*}
y_{ij} = f_{ij}\left(\bm\beta,\mathbf{u}_{i}\right) + \epsilon_{ij},
\end{equation*}
with $\mathbf{u}_{i} \sim \mathcal{N}\left(\mathbf{0}, \bm\Sigma\right)$ and $\epsilon_{ij} \sim \mathcal{N}\left(0, \sigma^2\right)$. First of all, conditionally on the random effects, both models impose a restriction on the error term \citep{powell}. However, the NLME model requires $\expect\left(\epsilon_{ij}|\mathbf{x}_{ij},\mathbf{u}_{i}\right) = 0$, while the AL-based specification of the error given in~\eqref{eq:5} leads to $Q_{\epsilon_{\tau,ij}|\mathbf{x}_{ij},\mathbf{u}_{i}}(\tau) = 0$ or, equivalently, $\Pr\left(\epsilon_{\tau,ij} < 0|\mathbf{x}_{ij},\mathbf{u}_{i}\right) = \tau$. Secondly, the fixed effects can be interpreted as the average value of the cluster-specific parameters, i.e. $\expect_{\mathbf{u}_{i}}\left(\bm\phi_{ij}\right)$, or as the regression parameters of the `zero-median' cluster, i.e. a cluster with a zero random-effect vector. However, the parameter $\bm\beta_{\tau}$ is allowed to vary with the quantile level $\tau$, while $\bm\beta$ in the NLME model is not (except for a location shift). Finally, in both approaches the variance-covariance matrix of the random effects gives a measure of the variability of $\mathbf{u}_{i}$ around $\expect_{\mathbf{u}_{i}}\left(\bm\phi_{ij}\right)$ but, again, estimates are allowed to differ by $\tau$ only for the quantile mixed-effects model.

In general, neither model~\eqref{eq:5} nor the NLME model provide fixed parameters that can be interpreted as, respectively, regression quantiles or regression means for the population. This is because random effects are allowed to enter nonlinearly in the model. (Similarly, several generalized linear mixed models with nonlinear link functions lack marginal interpretability \citep{ritz_spiegelman,gory}.) In contrast, the fixed effects of a linear (normal) mixed model remain the same after the random effects are integrated out, whereas, in general, this is not true for the fixed effects of the LQMMs of \cite{geraci2014}.

\subsection{Laplacian approximation}
\label{sec:3.2}

Let $\bm\Psi_{\tau} = \bm\Sigma_{\tau}/\sigma_{\tau}$ be the scaled variance-covariance matrix of the random effects and define $\bm\theta_{\tau}=\left(\bm\beta_{\tau}\tp,\bm\xi_{\tau}\tp\right)\tp$, where $\bm\xi_{\tau}$ an an unrestricted $m$-dimensional vector, $1 \leq m \leq q(q+1)/2$, of non-redundant parameters in $\bm\Psi_{\tau}$. Our goal is to maximize the marginal log-likelihood
\begin{align}\label{eq:6}
\nonumber \ell\left(\bm\theta_{\tau}; \mathbf{y}\right) & = C - \left(N + \frac{Mq}{2}\right)\log \sigma_{\tau} - \frac{M}{2} \log |\bm\Psi_{\tau}|\\
& \quad + \sum_{i = 1}^{M}\log\int_{\mathbb{R}^{q}}\exp\left\{-\frac{1}{\sigma_{\tau}}\sum_{j=1}^{n_{i}} \rho_\tau\left(y_{ij}-\mu_{\tau,ij}\right) - \frac{1}{2\sigma_{\tau}}\mathbf{u}_{i}\tp \bm{\Psi}_{\tau}^{-1}\mathbf{u}_{i}\right\} \rds \mathbf{u}_{i},
\end{align}
where $C = N \log \{\tau (1 - \tau)\} - Mq \log \sqrt{2\pi}$ and $\mu_{\tau,ij} = f_{ij}\left(\bm\beta_{\tau},\mathbf{u}_{i}\right)$.

First of all, we consider the following smooth approximation of $\rho_{\tau}$ \citep{madsen1993,chen2007}:
\begin{equation*}
\kappa_{\omega,\tau}(r)=\begin{cases}
r(\tau-1)-\frac{1}{2}(\tau-1)^2\omega & \text{if $r \leq (\tau-1)\omega$},\\
\frac{1}{2\omega}r^2& \text{if $(\tau - 1)\omega\leq r \leq \tau
\omega$},\\
r\tau-\frac{1}{2}\tau^2\omega & \text{if $r \geq \tau\omega$},
\end{cases}
\end{equation*}
where $r \in \mathbb{R}$ and $\omega > 0$ is a scalar ``tuning'' parameter. For $\omega \rightarrow 0$, we have that $\kappa_{\omega,\tau}(r) \rightarrow \rho_{\omega,\tau}(r)$. A similar approximation is given by \cite{muggeo_etal} who claimed that their method provides a better approximation than \citeauthor{chen2007}'s (\citeyear{chen2007}) algorithm. However, no analytical evidence was provided in their paper to support such claim. This point might offer scope for additional investigation but, here, it represents a secondary issue and will not be discussed any further.

Let $\mathbf{r}_{i} = (r_{i1},\ldots,r_{in_{i}})\tp$ be the vector of residuals $r_{ij} = y_{ij} - f\left(\bm\phi_{\tau,ij},\mathbf{x}_{ij}\right)$, $j = 1,\dots,n_{i}$, for the $i$th cluster, and define the corresponding sign vector $\mathbf{s}_{i} = (s_{i1},\ldots,s_{in_{i}})\tp$ with
\begin{equation}\label{eq:7}
s_{ij} =
\begin{cases}
-1 & \text{if $r_{ij} \leq (\tau-1)\omega$},\\
0 & \text{if $(\tau - 1)\omega < r_{ij} < \tau
\omega$},\\
1 &  \text{if $r_{ij} \geq \tau\omega$}.
\end{cases}
\end{equation}
(Note that the notation above has been simplified since the $r_{ij}$'s as well as the $s_{ij}$'s should be written as functions of the $\bm\phi_{\tau,ij}$'s.) Then we have
\begin{equation}\label{eq:8}
\sum_{j=1}^{n_{i}} \kappa_{\omega,\tau}(r_{ij}) = \frac{1}{2}\left( \frac{1}{\omega} \mathbf{r}_{i}\tp \mathbf{A}_{i} \mathbf{r}_{i} + \mathbf{b}_{i}\tp\mathbf{r}_{i} + \mathbf{c}_{i}\tp \mathbf{1}_{n_{i}}\right),
\end{equation}
where $\mathbf{A}_{i}$ is an $n_{i} \times n_{i}$ diagonal matrix with diagonal elements $\left\{\mathbf{A}_{i}\right\}_{jj} = 1 - s_{ij}^{2}$, $\mathbf{b}_{i}$ and $\mathbf{c}_{i}$ are two $n_{i} \times 1$ vectors with elements
\[
b_{ij} = s_{ij}((2\tau - 1)s_{ij} + 1)
\]
and
\[
c_{ij} = \frac{1}{2}\left\{(1 - 2\tau)\omega s_{ij} - (1 - 2\tau + 2\tau^2)\omega s_{ij}^2\right\},
\]
respectively.

\vskip .2cm

We now define the function
\begin{equation}\label{eq:9}
h\left(\bm\theta_{\tau}, \mathbf{y}_i, \mathbf{u}_i\right)=
\frac{1}{\omega} \mathbf{r}_{i}\tp \mathbf{A}_{i} \mathbf{r}_{i} + \mathbf{b}_{i}\tp\mathbf{r}_{i} + \mathbf{c}_{i}\tp \mathbf{1}_{n_{i}}
+ \mathbf{u}_{i}\tp\bm\Psi_{\tau}^{-1}\mathbf{u}_{i},
\end{equation}
which is akin to a regularized, nonlinear, weighted least-squares loss function. The gradient of $h$ with respect to $\mathbf{u}_i$ is given by
\begin{equation}\label{eq:10}
h'\left(\bm\theta_{\tau}, \mathbf{y}_i, \mathbf{u}_i\right) = -\mathbf{J}_{i}(\mathbf{u}_i)\tp \left[\frac{2}{\omega}\mathbf{A}_{i}\left\{\mathbf{y}_{i}-\mathbf{f}_{i}\left(\bm\beta_{\tau},\mathbf{u}_{i}\right)\right\} + \mathbf{b}_{i}\right] + 2\bm\Psi_{\tau}^{-1}\mathbf{u}_{i},
\end{equation}
where $\mathbf{f}_{i} = \left(f_{i1}\left(\bm\beta_{\tau}, \mathbf{u}_i\right), \ldots, f_{in_{i}}\left(\bm\beta_{\tau}, \mathbf{u}_i\right)\right)\tp$ and $\mathbf{J}_{i}(\mathbf{u}_i) = \partial \mathbf{f}_{i}\left(\bm\beta_{\tau}, \mathbf{u}_i\right)/\partial \mathbf{u}_{i}\tp$, while the Hessian is given by
\begin{align}\label{eq:11}
\nonumber h''\left(\bm\theta_{\tau}, \mathbf{y}_i, \mathbf{u}_i\right) &= \sum_{j = 1}^{n_{i}} \left\{- \frac{2}{\omega} \left(1 - s_{ij}^{2}\right)r_{ij}- b_{ij}\right\}\frac{\partial^{2} f_{ij}\left(\bm\beta_{\tau}, \mathbf{u}_i\right)}{\partial \mathbf{u}_{i}\partial \mathbf{u}_{i}\tp}\\
& \quad + \sum_{j = 1}^{n_{i}} \frac{\partial f_{ij}\left(\bm\beta_{\tau}, \mathbf{u}_i\right)}{\partial \mathbf{u}_{i}}\frac{\partial f_{ij}\left(\bm\beta_{\tau}, \mathbf{u}_i\right)}{\partial \mathbf{u}_{i}\tp} +  2\bm\Psi_{\tau}^{-1},
\end{align}
Moreover, let
\begin{equation}\label{eq:12}
\widehat{\mathbf{u}}_{i} = \argmin_{\mathbf{u}_{i}} h\left(\bm\theta_{\tau}, \mathbf{y}_i, \mathbf{u}_i\right)
\end{equation}
be the conditional mode of $\mathbf{u}_{i}$. For a given value of $\omega$, this can be obtained by means of penalized least-squares.

A second-order approximation of $h$ around $\widehat{\mathbf{u}}_{i}$ is given by
\begin{equation*}
h\left(\bm\theta_{\tau}, \mathbf{y}_i, \mathbf{u}_i\right) \simeq
h_{i} +
\dot{\mathbf{h}}_{i}\tp \left(\mathbf{u}_{i}-\widehat{\mathbf{u}}_i\right) +
\left(\mathbf{u}_{i}-\widehat{\mathbf{u}}_i\right)\tp \ddot{\mathbf{H}}_{i} \left(\mathbf{u}_{i}-\widehat{\mathbf{u}}_i\right),
\end{equation*}
where $h_{i}\equiv h\left(\bm\theta_{\tau}, \mathbf{y}_i, \widehat{\mathbf{u}}_i\right)$, $\dot{\mathbf{h}}_{i}\equiv
h'\left(\bm\theta_{\tau}, \mathbf{y}_i, \widehat{\mathbf{u}}_i\right)$, and $\ddot{\mathbf{H}}_{i}\equiv h''\left(\bm\theta_{\tau}, \mathbf{y}_i, \widehat{\mathbf{u}}_i\right)/2$. Since $\dot{\mathbf{h}}_{i}$ is zero at $\mathbf{u}_i = \widehat{\mathbf{u}}_i$, we have the following Laplacian approximation of the log-likelihood
\begin{align}\label{eq:13}
\nonumber \ell_{\mathrm{LA}}\left(\bm\theta_{\tau}; \mathbf{y}\right) & =N \log \left\{\frac{\tau (1 - \tau)}{\sigma_{\tau}}\right\} -\frac{M}{2} \log |\bm\Psi_{\tau}| -\frac{1}{2\sigma_{\tau}}\sum_{i = 1}^{M}h_{i} + \sum_{i=1}^{M} \log \int_{\mathbb{R}^{q}} (2\pi \sigma_{\tau})^{-q/2}\\
\nonumber &\quad \times \exp\left\{-\frac{1}{2\sigma_{\tau}} \left(\mathbf{u}_{i}-\widehat{\mathbf{u}}_i\right)\tp \ddot{\mathbf{H}}_{i} \left(\mathbf{u}_{i}-\widehat{\mathbf{u}}_i\right) \right\}\rds \mathbf{u}_{i}\\
& = N\log\left\{\frac{\tau (1 - \tau)}{\sigma_{\tau}}\right\} -\frac{1}{2} \left\{\sum_{i=1}^{M} \log |\bm\Psi_{\tau}\ddot{\mathbf{H}}_{i}| + \sigma_{\tau}^{-1}\sum_{i = 1}^{M}h_{i}\right\}.
\end{align}
If we ignore the negligible contribution of the first term in expression~\eqref{eq:11} \citep{pinheiro1995}, then only the first-order partial derivatives of $f$ are required to compute \eqref{eq:13}.

Since $\widehat{\mathbf{u}}_i$ does not depend on $\sigma$, the log-likelihood $\ell_{\mathrm{LA}}$ can be profiled on $\sigma$ leading to
\begin{equation}\label{eq:14}
\ell_{\mathrm{LA}_{p}}\left(\bm\theta_{\tau}; \mathbf{y}\right) = N\left[\log\left\{\frac{\tau (1 - \tau)}{\hat{\sigma}_{\tau}}\right\}-1\right] -\frac{1}{2} \sum_{i=1}^{M} \log |\bm\Psi_{\tau}\ddot{\mathbf{H}}_{i}|,
\end{equation}
where $\hat{\sigma}_{\tau} = (2N)^{-1}\sum_{i}^{M} h_{i}$.

Estimation of the parameters can be carried out iteratively. A pseudo-code of the algorithm is given in Appendix. The algorithm requires setting the starting value of $\bm\theta_{\tau}$ and $\sigma_{\tau}$, the tuning parameter $\omega$, the tolerance for the change in the log-likelihood, and the maximum number of iterations. Moreover, the modes of the random effects can be obtained by minimizing the objective function of the penalized least-squares problem using a Gauss-Newton method. Let $\bm\Delta_{\tau}$ be the relative precision factor such that $\bm\Psi_{\tau}^{-1} = \bm\Delta_{\tau}\tp \bm\Delta_{\tau}$ \citep{pinheiro_bates}. Then the function in~\eqref{eq:9} can be rewritten as
\begin{align}\label{eq:15}
\nonumber h\left(\bm\theta_{\tau}, \mathbf{y}_i, \mathbf{u}_i\right) &= \| \mathbf{A}^{1/2}_{i} \mathbf{r}_{i} \|^{2}/\omega + \mathbf{b}_{i}\tp\mathbf{r}_{i} + \mathbf{c}_{i}\tp \mathbf{1}_{n_{i}}
+ \| \bm\Delta_{\tau} \mathbf{u}_{i} \|^{2}\\
& =  \| \tilde{\mathbf{y}}_{i} - \tilde{\mathbf{f}}_{i} \|^{2} + \mathbf{b}_{i}\tp\left(\mathbf{y}_i - \mathbf{f}_{i}\right) + \mathbf{c}_{i}\tp \mathbf{1}_{n_{i}},
\end{align}
where
\begin{equation*}
\tilde{\mathbf{y}}_{i} = \left[
\begin{array}{c}
  \tilde{\mathbf{A}}_{i}\mathbf{y}_{i} \\
  \mathbf{0}
\end{array}
\right], \qquad
\tilde{\mathbf{f}}_{i} = \left[
\begin{array}{c}
  \tilde{\mathbf{A}}_{i}\mathbf{f}_{i} \\
  \bm\Delta_{\tau}\mathbf{u}_{i}
\end{array}
\right], \qquad
\tilde{\mathbf{A}}_{i} = \dfrac{1}{\sqrt{\omega}} \mathbf{A}^{1/2}_{i}.
\end{equation*}

When using the asymmetric Laplace as pseudo-likelihood, inference must be restrained to point estimation \citep[see for example][]{yang2016}. Standard errors for non-random parameters can be calculated using block bootstrap, although this increases the computational cost. Bootstrap confidence intervals have been shown to have good coverage in LQMMs \citep{geraci2014}.

\section{Simulation study}
\label{sec:4}
In this section, we perform a simulation study to evaluate the proposed methods. We start from a setting similar to the one used in \cite{pinheiro1995}, which is ideal for normal NLME models, and then investigate scenarios more apposite for NLQMM.

In the first scenario, we simulated the data from the following logistic model
\begin{equation}
\label{eq:16}
y_{ij} = \frac{\beta_{1} - \beta_{4} + u_{1i}}{1 + \exp\{(\beta_{2} + u_{2i} - x_{ij})/\beta_{3}\}} + (\beta_{4} + \epsilon_{ij}),
\end{equation}
where $\bm\beta = (70, 10, 3, 10)\tp$, $\mathbf{u}_{i} = (u_{1i},u_{2i})\tp \sim \mathcal{N}\left(\mathbf{0},\bm\Sigma\right)$, $x_{ij} \sim \mathcal{U}(0,20)$, and $\epsilon_{ij} \sim \mathcal{N}\left(0,1\right)$. The random effects are thus associated with the asymptotes ($\beta_{1}$ and $\beta_{4}$) and the sigmoid's midpoint ($\beta_{2}$). Their variance-covariance matrix was defined as
\begin{equation*}
\bm\Sigma =
\left[
  \begin{array}{cc}
    4 & -2 \\
    -2 & 5 \\
  \end{array}
\right].
\end{equation*}

In the second scenario, we used the same model \eqref{eq:16}, but we sampled the errors from a standardized chi-squared distribution with $3$ degrees of freedom, i.e. $\epsilon_{ij} \sim \chi_{3}^2/\sqrt{6}$.

In the third scenario, we slightly changed model \eqref{eq:16} and used
\begin{equation}
\label{eq:17}
y_{ij} = \frac{(\beta_{1} - \beta_{4})}{1 + \exp\left\{(\beta_{2} + u_{i} - x_{ij} - 0.5x_{ij}\epsilon_{ij})/\beta_{3}\right\}} + \beta_{4},
\end{equation}
where $\bm\beta = (1, 4, 1, 0)\tp$, $x_{ij} \sim \mathcal{U}(0,5)$, $u_{i} \sim \mathcal{N}\left(0,0.1\right)$, and $\epsilon_{ij} \sim \chi_{3}^2/\sqrt{60}$. Note that the error is skewed as in the second scenario but now operates within the exponential function. In this heteroscedastic model, there is only one random effect associated with the sigmoid's midpoint.

In the fourth and last scenario, we used the biexponential model
\begin{align}
\label{eq:18}
\nonumber y_{ij} = &  \left(\beta_{1} + u_{1i}\right)\exp\left\{-\exp\left(\beta_{2} + u_{2i}\right)x_{ij}\right\}\\
&+ \left(\beta_{3} + u_{3i}\right)\exp\left\{-\exp\left(\beta_{4} + u_{4i}\right)x_{ij}\right\} + (1 - x_{ij}/8)\epsilon_{ij},
\end{align}
where $\mathbf{u}_{i} = (u_{1i},u_{2i},u_{3i},u_{4i})\tp \sim \mathcal{N}\left(\mathbf{0},\bm\Sigma\right)$, $x_{ij} \sim \mathcal{U}(0,8)$, and $\epsilon_{ij} \sim \mathcal{N}\left(0,0.1\right)$, with parameters $\bm\beta = (2, 0.8, 0.4, -1.5)\tp$ and $\bm\Sigma = 0.1\mathbf{I}_{4}$.

\begin{figure}[h!]
\centering
\includegraphics[scale=0.6]{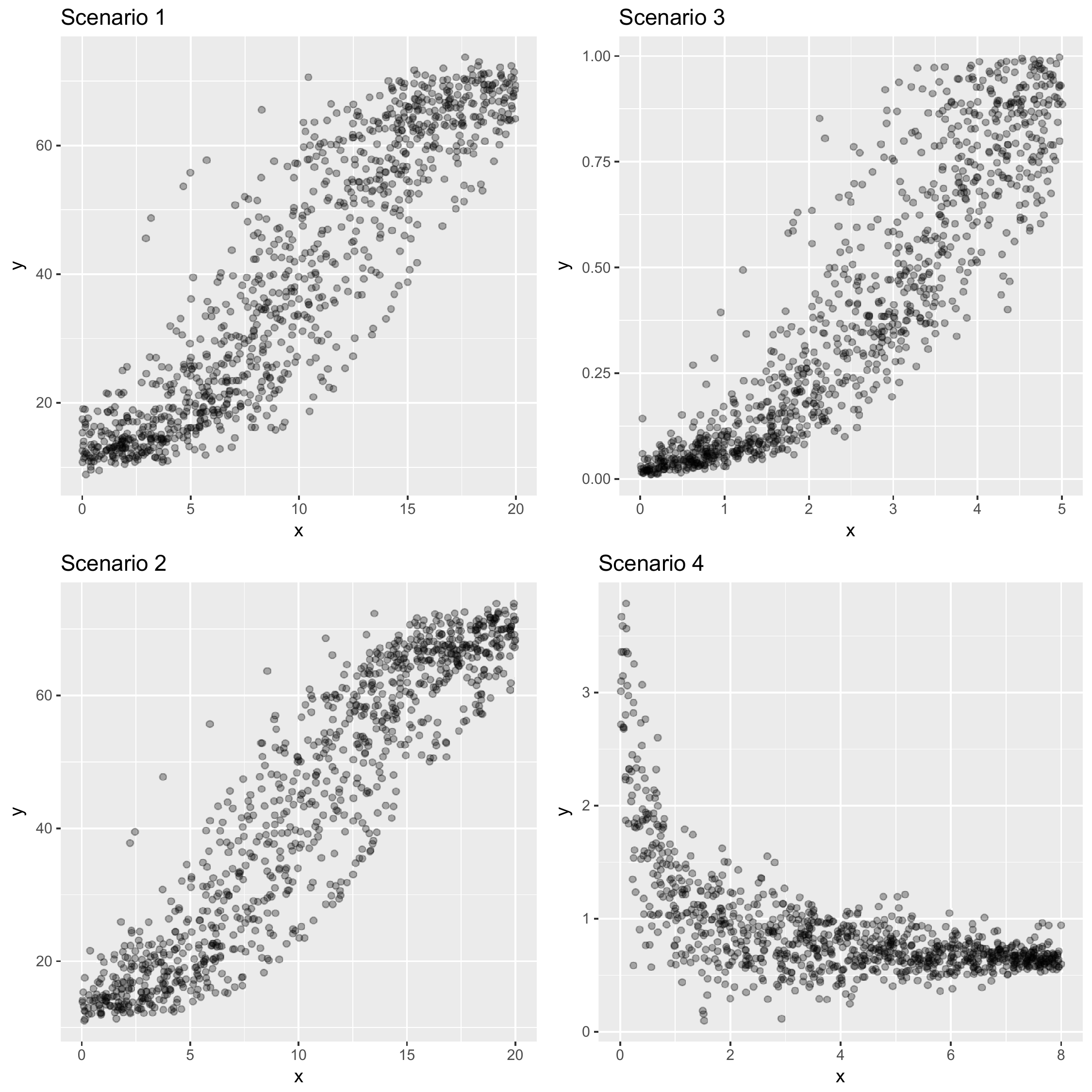}
\caption{Examples of data generated from the logistic (scenarios 1-3) and the biexponential (scenario 4) models.}
\label{fig:1}
\end{figure}

In all scenarios, we used $M = 100$, $n_{i} = 10$, $i = 1, \ldots, M$. Instances of replications are shown in Figure~\ref{fig:1}. For data sampled from models \eqref{eq:16} and \eqref{eq:17}, we fitted mixed-effects logistic quantile functions with parameter $\bm\phi_{\tau,ij} = \mathbf{F}_{ij}\bm\beta_{\tau} + \mathbf{G}_{ij}\mathbf{u}_{i}$, where
\[
\mathbf{F}_{ij} =
\left[
  \begin{array}{rrrr}
    1 & 0 & 0 & -1\\
    0 & 1 & 0 & 0\\
    0 & 0 & 1 & 0\\
    0 & 0 & 0 & 1
  \end{array}
\right]
\]
in the first 3 scenarios, \[
\mathbf{G}_{ij} =
\left[
  \begin{array}{l}
    \mathbf{I}_{2}\\
    \mathbf{O}_{2 \times 2}
  \end{array}
\right]\] in the first and second scenarios, and $\mathbf{G}_{ij} = (0, 1, 0, 0)\tp$ in the third scenario. For data sampled from model~\eqref{eq:18}, we fitted mixed-effects biexponential quantile functions with parameter $\bm\phi_{\tau,ij} = \mathbf{F}_{ij}\bm\beta_{\tau} + \mathbf{G}_{ij}\mathbf{u}_{i}$, where $\mathbf{F}_{ij} = \mathbf{G}_{ij} = \mathbf{I}_{4}$.

For each scenario, we replicated $R = 500$ datasets and fitted NLQMMs at three quantile levels using $\tau \in \{0.1,0.5,0.9\}$. Estimation was carried out by following the algorithm as described in Appendix. An attempt to maximize the approximated Laplacian log-likelihood \eqref{eq:13} was made by using the Broyden-Fletcher-Goldfarb-Shanno (BFGS) algorithm in the first instance. Upon failure of the BFGS algorithm during any iteration of the main estimation algorithm, the latter was started again and a new attempt to maximize \eqref{eq:13} was made by using the Nelder-Mead algorithm. The maximum number of iterations was set to 500, while the tolerance for the relative change in the log-likelihood was set to $10^{-4}$. Between two successive iterations, the tuning parameter $\omega$ was multiplied by the factor $\gamma = 0.2$. All the parameters of the optimization algorithms in \texttt{optim} and \texttt{nlm} were set at their default values. Computations were performed using the \texttt{R} environment for statistical computing and graphics \citep{R} version 3.3.2 on a desktop computer with a 3.60GHz quad core i7-4790 processor and 16 gigabytes of RAM.

Before we proceed with the analysis of the results, it is important to note that, in general, the nonlinearity of the models along with the presence of the random effects pose a difficulty for establishing the `true' value of $\bm\beta_{\tau}$ for quantiles other than the median \cite[see for example the simulation study in][]{karlsson2008}, even when the errors are normal. For example, in the logistic model not only the asymptotes $\beta_{\tau,1}$ and $\beta_{\tau,4}$, but also the midpoint $\beta_{\tau,2}$ and the scale $\beta_{\tau,3}$ change with $\tau$ in a rather complicated way. (An exception is given by model~\eqref{eq:17} for which the lower and upper asymptotes ($\beta_{\tau,4}$ and $\beta_{\tau,1}$, respectively) do not change with $\tau$.) We find solace in observing that such limitation brings out one of the advantages of quantile-based over moment-based modeling, since direct estimation of conditional quantiles does not require nontrivial manipulation of nonlinear relationships \citep[][p.435]{demidenko_2013}. As a reference, we can consider the corresponding results from standard nonlinear quantile regression (NLRQ) \citep{koenker_park} under the assumption of independent observations. Similarity of the magnitude and direction of the estimates would support the interpretation of $\bm\beta_{\tau}$ as regression parameters of the `zero-median' cluster, while comparing the variability of the estimates from NLQMM and NLRQ would inform us on whether accounting for clustering provides a gain in efficiency.

\begin{figure}
\centering
\includegraphics[scale=0.6]{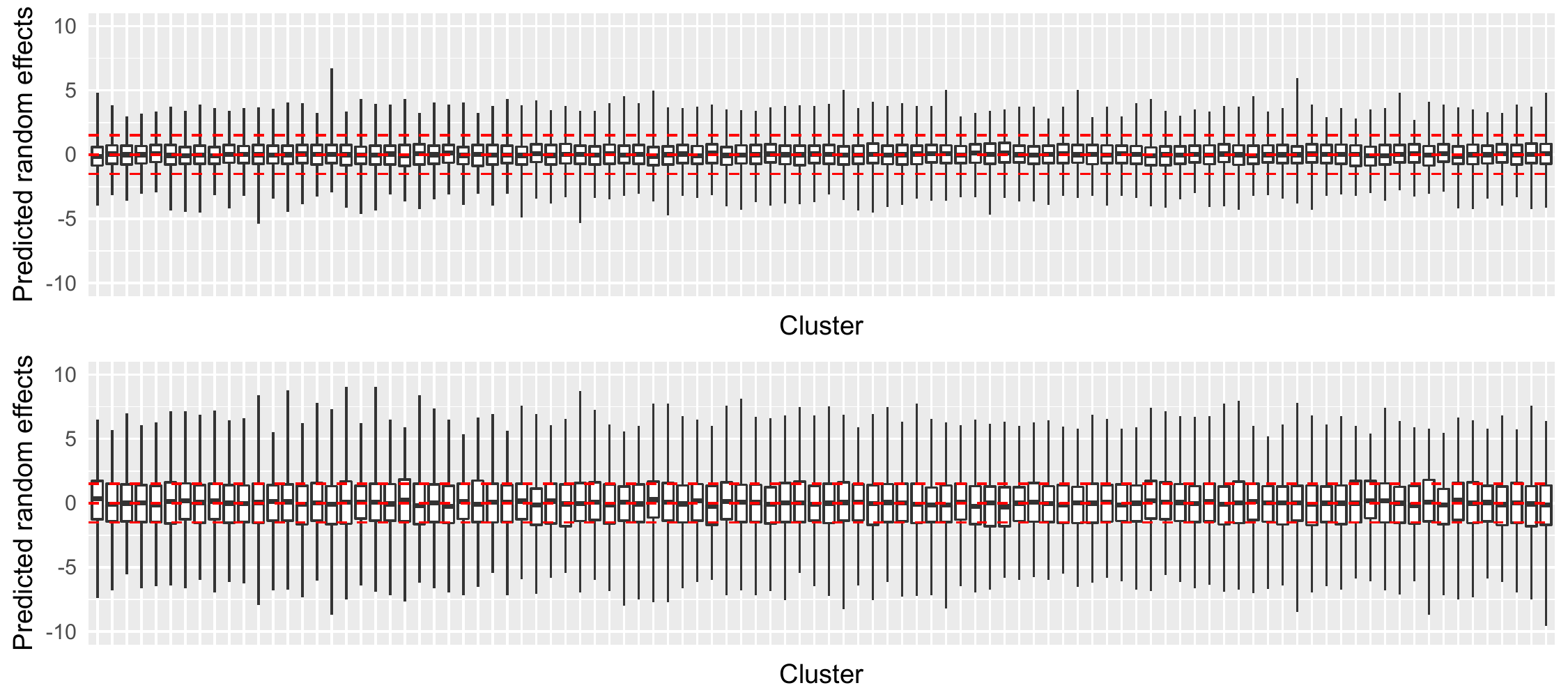}
\caption{Boxplots of random effects predicted from the median model for 100 clusters based on 500 replications from the first scenario. The dashed red lines mark the 25th, 50th, and 75th percentiles of the true distribution of the random effects.}
\label{fig:ex1}
\end{figure}

\begin{figure}
\centering
\includegraphics[scale=0.6]{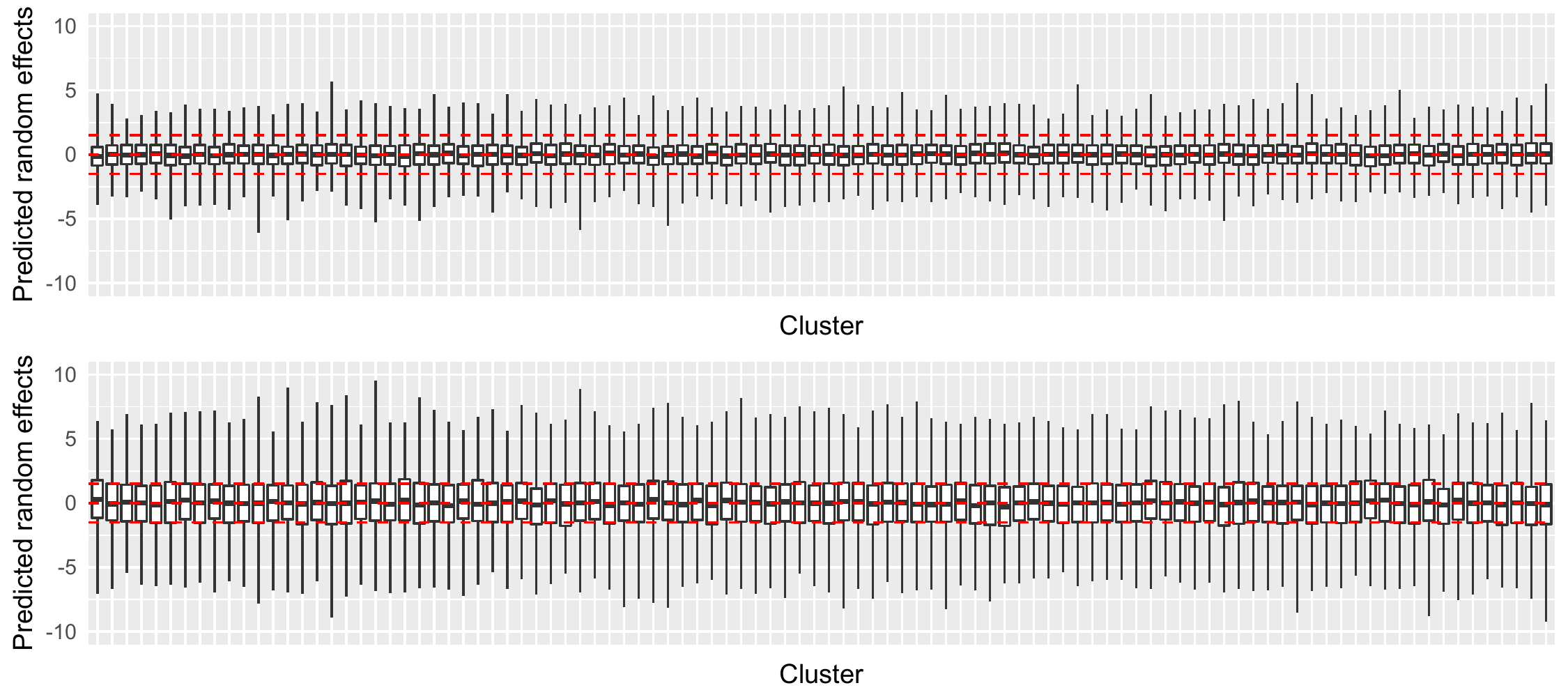}
\caption{Boxplots of random effects predicted from the median model for 100 clusters based on 500 replications from the second scenario. The dashed red lines mark the 25th, 50th, and 75th percentiles of the true distribution of the random effects.}
\label{fig:ex2}
\end{figure}

\begin{figure}
\centering
\includegraphics[scale=0.6]{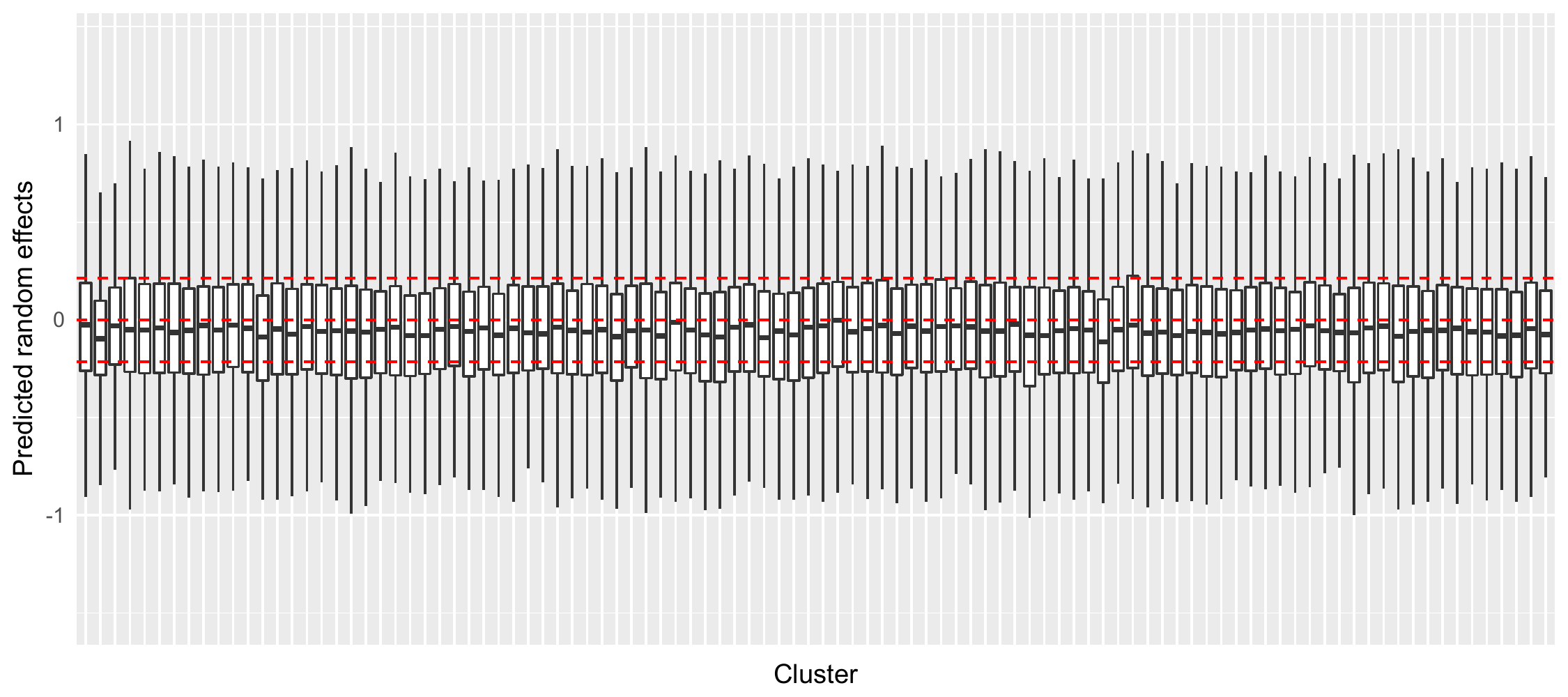}
\caption{Boxplots of random effects predicted from the median model for 100 clusters based on 500 replications from the third scenario. The dashed red lines mark the 25th, 50th, and 75th percentiles of the true distribution of the random effects.}
\label{fig:ex3}
\end{figure}

\begin{figure}
\centering
\includegraphics[scale=0.6]{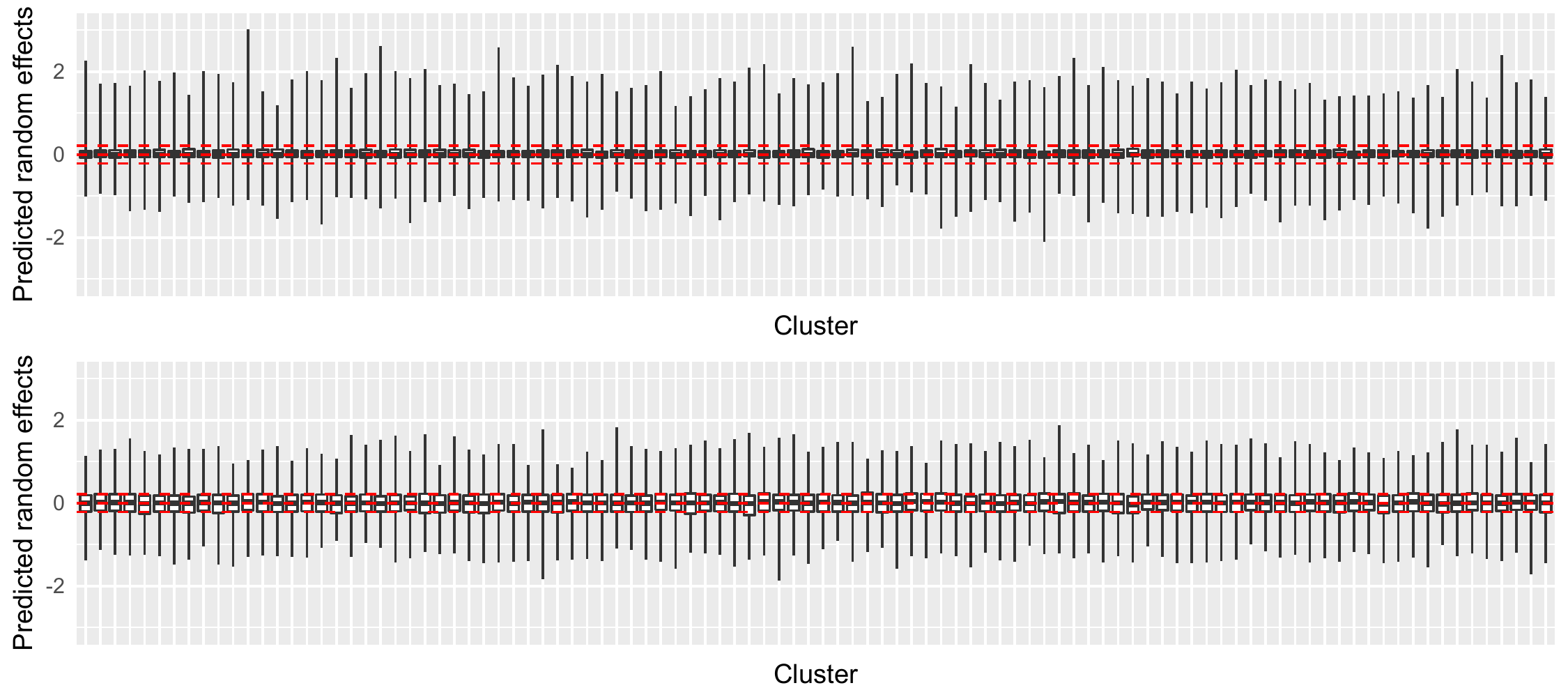}
\caption{Boxplots of random effects predicted from the median model for 100 clusters based on 500 replications from the fourth scenario. The dashed red lines mark the 25th, 50th, and 75th percentiles of the true distribution of the random effects.}
\label{fig:ex4}
\end{figure}

The average estimates $\hat{\bm\beta}_{\tau}$ and standard deviations of the estimates are reported in Tables~\ref{tab:1}-\ref{tab:4}. In summary, NLQMM estimates were close to NLRQ estimates in all scenarios. The variability of the estimates from NLQMM was either lower or close to that of the estimates from NLQR. Of all the results, perhaps those related to the quantile $0.9$ in the third scenario (Table~\ref{tab:3}) deserve more discussion. Both NLQMM and NLRQ clearly failed to provide meaningful estimates of the parameters. This is due to the fact that the range of the simulated values for $x$ was not wide enough to correctly estimate the upper asymptote at upper quantiles. This observation may have a particular relevance when modeling reference growth curves. Further, the estimated variance-covariance parameters (Table~\ref{tab:ex4}) and the predicted random effects obtained from \eqref{eq:12} (Figures~\ref{fig:ex1}-\ref{fig:ex4}) were, in general, consistent with the parameters of the true distribution of the random effects, although, as noted before, direct comparisons are not straightforward.

We now provide basic details about the algorithm's performance and a few recommendations. On average, it took about 7 iterations (approximately 35 seconds) to fit one model for the quantiles $0.1$ or $0.9$, and about 6 iterations (approximately 20 seconds) for the median in the first two scenarios. In the third scenario it took between $2$ and $7$ iterations (approximately $20$ seconds on average) to fit one model for any of the three quantile levels. In the fourth scenario, the algorithm needed a similar number of iterations as in the first two scenarios but the time to convergence was, on average, twice as long. This means that, within each iteration, the number of function evaluations required by \texttt{optim} to fit the more complex biexponential model was greater than that needed to fit the logistic model. In the first two scenarios, the median value of the factor $\gamma$ at the last iteration was about $2.0 \times 10^{-3}$ for all considered quantiles. In the third scenario, it was less than $4.5 \times 10^{-5}$ for the tail quantiles and $0.5$ for the median. In the fourth scenario, it was less than $4.5 \times 10^{-5}$ for all considered quantiles.

In a separate analysis (results not shown), the average number of iterations to convergence increased to at least 10 when $\gamma$ was increased to $0.5$. In contrast, the algorithm converged too quickly to smaller values of the log-likelihood when setting $\gamma$ to less than 0.2. This was to be expected since $\gamma$ controls the speed at which the smoothing parameter $\omega$ approaches zero. As in \cite{chen2007}, we recommend using $\gamma = 0.5$ in most situations.

Further, in the first three scenarios the average number of iterations and the values of the estimates were not particularly sensitive to the specific algorithm used for optimizing the log-likelihood, although the BFGS algorithm did fail to converge in about $20\%$ of the replications, more often when estimating tail quantiles ($28\%$) rather than when estimating the median ($12\%$). In contrast, BFGS never failed to converge in the fourth scenario. We then ran a separate analysis (results not shown) in which biexponential models were fitted exclusively using Nelder-Mead. For $\tau = 0.1$, estimates were unreasonable. We recommend using BFGS as default optimization algorithm.

\begin{table}[ht]
\caption{Estimates of the fixed effects from nonlinear quantile mixed-effects regression (NLQMM) and from nonlinear quantile regression (NLRQ) with $\tau \in \{0.1,0.5,0.9\}$ for the first scenario. The estimates are averaged over 500 replications and the standard deviations are reported in brackets.}
\centering
\begin{tabular}{lrrrr}
  \hline
 & \multicolumn{1}{c}{$\hat{\beta}_1$} & \multicolumn{1}{c}{$\hat{\beta}_2$} & \multicolumn{1}{c}{$\hat{\beta}_3$} & \multicolumn{1}{c}{$\hat{\beta}_4$} \\
  \hline
\multicolumn{5}{l}{NLQMM}\\
  \hline
  $\tau = 0.1$ & 68.00 (0.73) & 12.66 (0.40) & 3.06 (0.10) & 8.79 (0.34) \\
  $\tau = 0.5$ & 70.23 (0.38) & 9.99 (0.28) & 3.04 (0.05) & 9.70 (0.22) \\
  $\tau = 0.9$ & 73.47 (0.56) & 7.28 (0.38) & 3.15 (0.10) & 9.76 (0.52) \\
  \hline
\multicolumn{5}{l}{NLRQ}\\
  \hline
  $\tau = 0.1$ & 68.47 (2.76) & 12.77 (0.58) & 3.04 (0.27) & 9.54 (0.53) \\
  $\tau = 0.5$ & 69.79 (0.92) & 9.99 (0.33) & 3.00 (0.18) & 10.11 (0.71) \\
  $\tau = 0.9$ & 72.26 (0.85) & 7.19 (0.53) & 3.11 (0.32) & 9.53 (2.75) \\
\hline
\end{tabular}
\label{tab:1}
\end{table}

\begin{table}[ht]
\caption{Estimates of the fixed effects from nonlinear quantile mixed-effects regression (NLQMM) and from nonlinear quantile regression (NLRQ) with $\tau \in \{0.1,0.5,0.9\}$ for the second scenario. The estimates are averaged over 500 replications and the standard deviations are reported in brackets.}
\centering
\begin{tabular}{lrrrr}
  \hline
 & \multicolumn{1}{c}{$\hat{\beta}_1$} & \multicolumn{1}{c}{$\hat{\beta}_2$} & \multicolumn{1}{c}{$\hat{\beta}_3$} & \multicolumn{1}{c}{$\hat{\beta}_4$}\\
   \hline
\multicolumn{5}{l}{NLQMM}\\
  \hline
  $\tau = 0.1$ & 69.28 (0.68) & 12.70 (0.40) & 3.05 (0.08) & 10.25 (0.22) \\
  $\tau = 0.5$ & 71.39 (0.35) & 9.97 (0.27) & 3.05 (0.05) & 10.64 (0.19) \\
  $\tau = 0.9$ & 74.67 (0.55) & 7.26 (0.38) & 3.15 (0.10) & 10.98 (0.58) \\
  \hline
\multicolumn{5}{l}{NLRQ}\\
  \hline
  $\tau = 0.1$ & 69.67 (2.53) & 12.77 (0.55) & 3.03 (0.24) & 10.80 (0.46) \\
  $\tau = 0.5$ & 71.03 (0.90) & 9.98 (0.31) & 3.01 (0.18) & 11.25 (0.72) \\
  $\tau = 0.9$ & 73.49 (0.82) & 7.17 (0.54) & 3.11 (0.32) & 10.73 (2.86) \\
   \hline
\end{tabular}
\label{tab:2}
\end{table}

\begin{table}[ht]
\caption{Estimates of the fixed effects from nonlinear quantile mixed-effects regression (NLQMM) and from nonlinear quantile regression (NLRQ) with $\tau \in \{0.1,0.5,0.9\}$ for the third scenario. The estimates are averaged over 500 replications and the standard deviations are reported in brackets.}
\centering
\begin{tabular}{lrrrr}
  \hline
 & \multicolumn{1}{c}{$\hat{\beta}_1$} & \multicolumn{1}{c}{$\hat{\beta}_2$} & \multicolumn{1}{c}{$\hat{\beta}_3$} & \multicolumn{1}{c}{$\hat{\beta}_4$}\\
   \hline
\multicolumn{5}{l}{NLQMM}\\
  \hline
  $\tau = 0.1$ & 0.95 (0.06) & 4.18 (0.20) & 0.92 (0.08) & 0.00 (0.00) \\
  $\tau = 0.5$ & 1.00 (0.03) & 3.19 (0.08) & 0.85 (0.05) & 0.00 (0.01) \\
  $\tau = 0.9$ & $-$0.25 (0.93) & 2.09 (1.34) & $-$1.16 (1.40) & 1.11 (0.36) \\
  \hline
\multicolumn{5}{l}{NLRQ}\\
  \hline
  $\tau = 0.1$ & 1.01 (0.03) & 3.96 (0.06) & 1.01 (0.02) & $-$0.00 (0.00) \\
  $\tau = 0.5$ & 1.00 (0.04) & 3.18 (0.10) & 0.87 (0.06) & $-$0.00 (0.01) \\
  $\tau = 0.9$ & $-$0.48 (0.57) & $-$1.24 (1.65) & $-$2.19 (0.98) & 0.40 (0.22) \\
   \hline
\end{tabular}
\label{tab:3}
\end{table}

\begin{table}[ht]
\caption{Estimates of the fixed effects from nonlinear quantile mixed-effects regression (NLQMM) and from nonlinear quantile regression (NLRQ) with $\tau \in \{0.1,0.5,0.9\}$ for the fourth scenario. The estimates are averaged over 500 replications and the standard deviations are reported in brackets.}
\centering
\begin{tabular}{lrrrr}
  \hline
 & \multicolumn{1}{c}{$\hat{\beta}_1$} & \multicolumn{1}{c}{$\hat{\beta}_2$} & \multicolumn{1}{c}{$\hat{\beta}_3$} & \multicolumn{1}{c}{$\hat{\beta}_4$} \\
  \hline
  \multicolumn{5}{l}{NLQMM}\\
\hline
 $\tau = 0.1$ & 1.93 (0.19) & 1.06 (0.12) & 0.46 (0.13) & $-\infty$ ($\infty$) \\
  $\tau = 0.5$ & 2.04 (0.15) & 0.72 (0.10) & 1.03 (0.11) & $-$3.19 (0.17) \\
  $\tau = 0.9$ & 2.05 (0.17) & 0.61 (0.15) & 1.82 (0.12) & $-$2.48 (0.11) \\
  \hline
\multicolumn{5}{l}{NLRQ}\\
  \hline
  $\tau = 0.1$ & 1.86 (0.26) & 0.98 (0.19) & 0.61 (0.14) & $-\infty$ ($\infty$) \\
  $\tau = 0.5$ & 2.04 (0.18) & 0.70 (0.14) & 1.01 (0.13) & $-$3.27 (0.24) \\
  $\tau = 0.9$ & 2.17 (0.21) & 0.61 (0.19) & 1.55 (0.14) & $-$2.32 (1.89) \\
\hline
\end{tabular}
\label{tab:4}
\end{table}

\begin{table}[ht]
\caption{Estimates of the variance-covariance parameters from nonlinear quantile mixed-effects regression with $\tau \in \{0.1,0.5,0.9\}$ for all scenarios. The estimates are averaged over 500 replications.}
\centering
\begin{tabular}{lrrr}
  \hline
 & $\tau = 0.1$ & $\tau = 0.5$ & $\tau = 0.9$ \\
  \hline
  \multicolumn{4}{l}{\textit{First scenario}}\\
  \hline
  $\sigma_{1}^2$ & 1.73 & 0.97 & 1.68 \\
  $\sigma_{12}$ & $-$3.30 & $-$1.83 & $-$3.19 \\
  $\sigma_{2}^2$ & 6.61 & 3.67 & 6.36 \\
  \hline
  \multicolumn{4}{l}{\textit{Second scenario}}\\
  \hline
  $\sigma_{1}^2$ & 1.76 & 0.99 & 1.73 \\
  $\sigma_{12}$ & $-$3.34 & $-$1.85 & $-$3.25 \\
  $\sigma_{2}^2$ & 6.62 & 3.65 & 6.38 \\
  \hline
  \multicolumn{4}{l}{\textit{Third scenario}}\\
  \hline
  $\sigma_{1}^2$ & 0.01 & 0.04 & 0.00 \\
  \hline
  \multicolumn{4}{l}{\textit{Fourth scenario}}\\
  \hline
  $\sigma_{1}^2$ & 0.16 & 0.22 & 0.04 \\
  $\sigma_{2}^2$ & 0.24 & 0.25 & 0.08 \\
  $\sigma_{3}^2$ & 0.16 & 0.18 & 0.17 \\
  $\sigma_{4}^2$ & 0.09 & 0.17 & 0.01 \\
   \hline
\end{tabular}
\label{tab:ex4}
\end{table}

\clearpage

\clearpage

\section{Applications}
\label{sec:5}
\subsection{Pharmacokinetics}
\label{sec:5.1}

\begin{figure}[b!]
\centering
\includegraphics[scale=0.6]{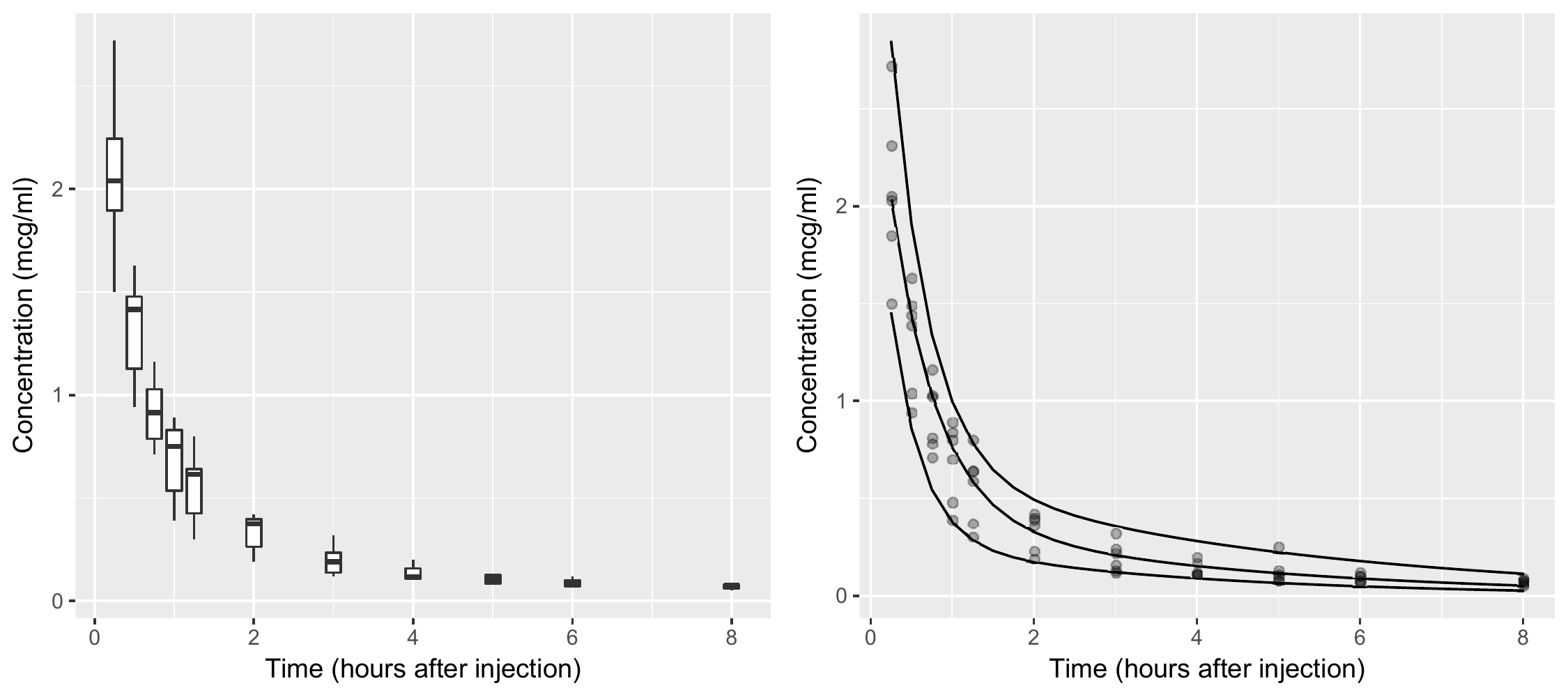}
\caption{Boxplots of indomethacin concentration by measurement occasion (left) and fitted biexponential curves at the 10th, 50th and 90th centiles of drug concentration conditional on time since injection (right).}
\label{fig:2}
\end{figure}

We begin with the analysis of a dataset taken from an old pharmacokinetics study \citep{kwan}, often used as a toy example in nonlinear regression modeling \citep{davidian_1995,pinheiro_bates}. The data consists of 11 measurements of plasma concentrations of indomethacin which was injected intravenously in 6 subjects. We used the biexponential model
\begin{align*}
Q_{y_{ij}|\mathbf{u}_{i}}(\tau) = & \left(\beta_{\tau,1} + u_{1i}\right)\exp\left\{-\exp\left(\beta_{\tau,2} + u_{2i}\right)t_{j}\right\} \\
& + \left(\beta_{\tau,3} + u_{3i}\right)\exp\left\{-\exp\left(\beta_{\tau,4}\right)t_{j}\right\},
\end{align*}
where $y_{ij}$ denotes the $j$th measurement of drug concentration (mcg/ml), $j = 1, \ldots, 11$, on the $i$th subject, $i = 1,\ldots, 6$, and $t_{j}$ is the time (hr) of the measurement since injection (given that the dataset is balanced, $t$ does not depend on $i$). We modeled the variance-covariance of the random effects using the diagonal matrix $\bm\Sigma_{\tau} = \bigoplus_{k=1}^{3}\sigma^{2}_{\tau,k}$ (variance components). Note that the regression model above includes 3 random effects, one for each of the first 3 fixed effects. In a separate analysis (results not shown), we found that the random effect associated with $\beta_{\tau,4}$, $\tau \in \{0.1,0.5,0.9\}$, had near-zero variance \citep[see also][p.283]{pinheiro_bates}.

In this two-compartment model, the first exponential term determines the initial, rapidly declining distribution phase of the drug. The elimination of the drug is the predominant process during the second phase and is primarily determined by the second exponential term. Besides the average rates at which the drug is distributed and then eliminated, it might be of interest to assess the change over time of concentrations that are higher or lower than the mean. The left plot of Figure~\ref{fig:2} shows the boxplots of indomethacin concentration at each measurement occasion. It appears that the scale and possibly even the shape of the distribution are changing over time. This would mean that the rates are not similar across the quantiles of the conditional distribution.

\begin{table}[ht]
\caption{Estimates of the fixed effects (standard errors) from three biexponential quantile mixed-effects models with $\tau \in \{0.1,0.5,0.9\}$ and from the normal nonlinear mixed-effects model (NLME) using the Indomethacin Data. Standard errors for quantile regression estimates are based on 200 bootstrap replications. Bold denotes statistically significant at the $5\%$ level.}
\centering
\begin{tabular}{lcccc}
  \hline
 & $\beta_1$ & $\beta_2$ & $\beta_3$ & $\beta_4$ \\
  \hline
$\tau = 0.1$ & \textbf{2.31} (0.48) & \textbf{0.99} (0.16) & \textbf{0.30} (0.13) & \textbf{$-$1.19} (0.57) \\
  $\tau = 0.5$ & \textbf{2.55} (0.28) & \textbf{0.58} (0.19) & \textbf{0.44} (0.17) & \textbf{$-$1.33} (0.23) \\
  $\tau = 0.9$ & \textbf{3.73} (0.52) & \textbf{0.75} (0.35) & \textbf{0.69} (0.34) & \textbf{$-$1.49} (0.37) \\
  NLME & \textbf{2.83} (0.26) & \textbf{0.77} (0.11) & \textbf{0.46} (0.11) & \textbf{$-$1.35} (0.23) \\
   \hline
\end{tabular}
\label{tab:5}
\end{table}

\begin{table}[ht]
\caption{Estimates of the variance components from three biexponential quantile mixed-effects models with $\tau \in \{0.1,0.5,0.9\}$ and from the normal nonlinear mixed-effects model (NLME) using the Indomethacin Data.}
\centering
\begin{tabular}{lccc}
  \hline
 & $\sigma^{2}_1$ & $\sigma^{2}_2$ & $\sigma^{2}_3$ \\
  \hline
$\tau = 0.1$ & 0.78 & 0.06 & 0.02 \\
  $\tau = 0.5$ & 0.59 & 0.08 & 0.02 \\
  $\tau = 0.9$ & 1.34 & 0.05 & 0.06 \\
  NLME & 0.33 & 0.03 & 0.01 \\
   \hline
\end{tabular}
\label{tab:6}
\end{table}

The fitted biexponential curves for $\tau \in \{0.1,0.5,0.9\}$ are given in the right plot of Figure~\ref{fig:2}, while estimates of the fixed effects and their standard errors are reported in Table~\ref{tab:5}. The average rate (NLME) at which the drug decreases during the distribution phase was comparable to that of the 90th centile. However, the decrease was about $20\%$ \emph{faster} at the lower 10th centile but about $20\%$ \emph{slower} at the median as compared to the mean. During the second phase, the rate of decrease was, again, greatest at the 10th centile. However, the average rate was similar to that of the median and greater than that of the 90th centile. One implication is that the distribution of the response becomes increasingly right-skewed as time passes. This is an advantage of NLQMM over NLME as the latter cannot obviously model changes in the shape of the distribution.

Finally, there was substantial heterogeneity among subjects regarding starting concentration levels during the distribution phase, especially at the 90th centile (Table~\ref{tab:6}).

\subsection{Growth curves}
\label{sec:5.1}
In this section, we analyze data on growth patterns of two genotypes of soybeans: Plant Introduction $\#416937$ (P), an experimental strain, and Forrest (F), a commercial variety \citep{davidian_1995}. The response variable is the average leaf weight of 6 plants chosen at random from each experimental plot and measured at approximately weekly intervals, between two and eleven weeks after planting. The experiment was carried out over three different planting years: 1988, 1989, and 1990. Eight plots were planted with each genotype in each planting year, giving a total of 48 plots in the study \citep{pinheiro_bates}.

\begin{figure}
\centering
\includegraphics[scale=0.6]{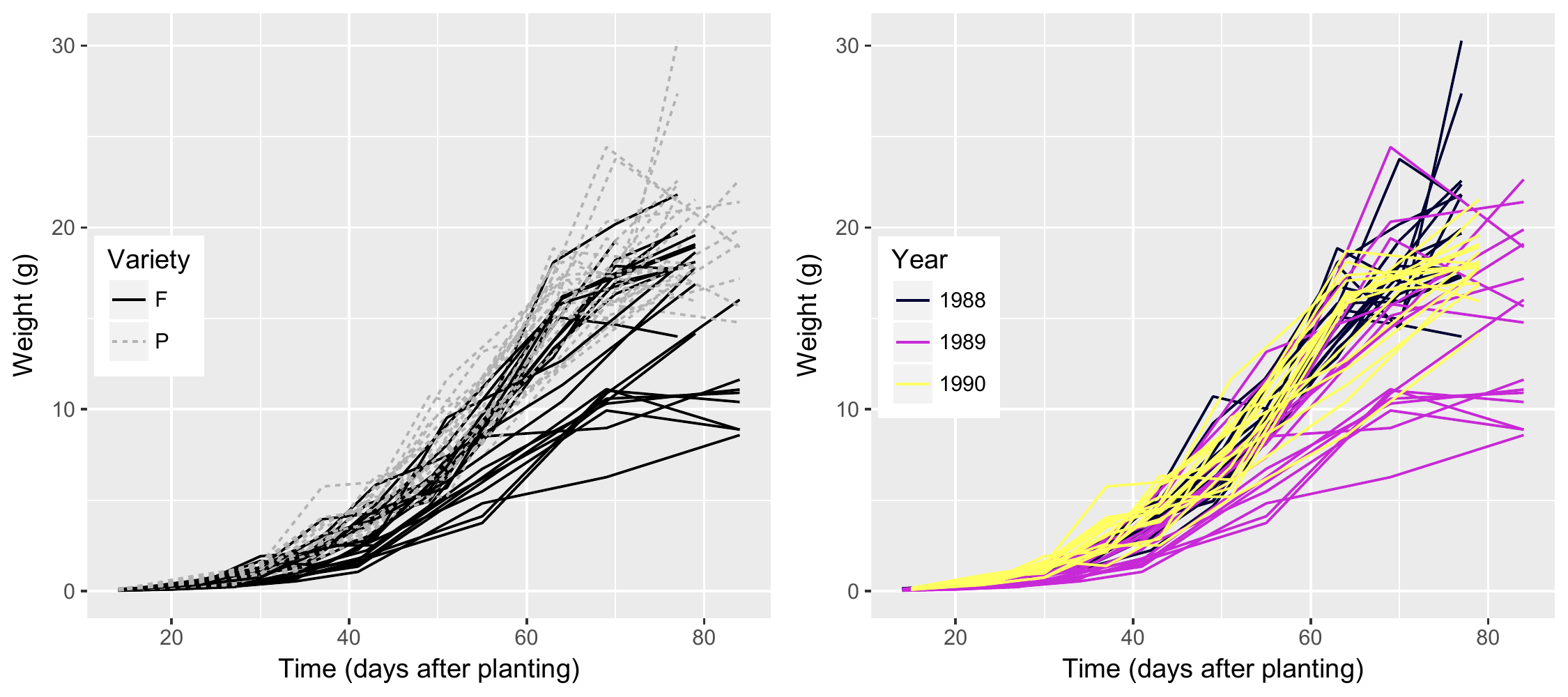}
\caption{Observed growth curves of soybean plants. Each line represents the average leaf weight per plant in each experimental plot. Curves are grouped by variety (left) or by year (right).}
\label{fig:3}
\end{figure}

Figure~\ref{fig:3} shows the temporal trajectories of the average leaf weight for individual plots. It is apparent that the experimental strain yielded heavier leaves that the F variety, at least on average. There also seem to be differences between planting years, with a wider spread of the curves in 1989. Given that previous analyses of these data focused on the average growth curves \citep{davidian_1995,pinheiro_bates}, we set out to investigate growth in the tails of the distribution. For our analysis, we used the same logistic model as that in \citet[][p.293]{pinheiro_bates} which was selected over a number of alternative models, that is
\begin{equation*}
Q_{y_{ij}|\mathbf{u}_{i}}(\tau) = \frac{\phi_{\tau,1ij}}{1 + \exp\{(\phi_{\tau,2ij} - t_{ij})/\phi_{\tau,3ij}\}},
\end{equation*}
where $y_{ij}$ denotes the average leaf weight (g) observed on occasion $j$, $j = 1, \ldots, n_{i}$, in the $i$th plot, $i = 1,\ldots, 48$, and $t_{ij}$ is the time (days) of the measurement after planting. The number of measurements per plot ranged between 8 and 10. The design matrices of the $3\times 1$ parameter $\bm\phi_{\tau,ij} = \mathbf{F}_{ij}\bm\beta_{\tau} + \mathbf{G}_{ij}u_{i}$ were given by

\begin{equation*}
\mathbf{F}_{ij} =
\left[
  \begin{array}{lllllllllllll}
    1 & x^{(89)}_{ij} & x^{(90)}_{ij} & x^{(P)}_{ij} & x^{(89)}_{ij}\cdot x^{(P)}_{ij} & x^{(90)}_{ij}\cdot x^{(P)}_{ij} & 0 & 0 & 0 & 0 & 0 & 0 & 0\\
    0 &0 & 0& 0& 0& 0& 1 & x^{(89)}_{ij} & x^{(90)}_{ij} & x^{(P)}_{ij} & 0&0 &0 \\
    0 &0 & 0& 0& 0& 0& 0 & 0 & 0 & 0 & 1 & x^{(89)}_{ij} & x^{(90)}_{ij}
  \end{array}
\right]
\end{equation*}
and $\mathbf{G}_{ij} = \left[\begin{array}{lll} 1 & 0 & 0 \end{array}\right]\tp$. Thus, $\bm\beta_{\tau}$ is a $13 \times 1$ vector. The covariates in the $F$ matrix are dummy variables for year of planting, $x^{(89)}$ and $x^{(90)}$, and genotype, $x^{(P)}$. The baseline is represented by year 1988 and variety F. The only random effect included in the model was associated with the asymptote.

\begin{figure}
\centering
\includegraphics[scale=0.45]{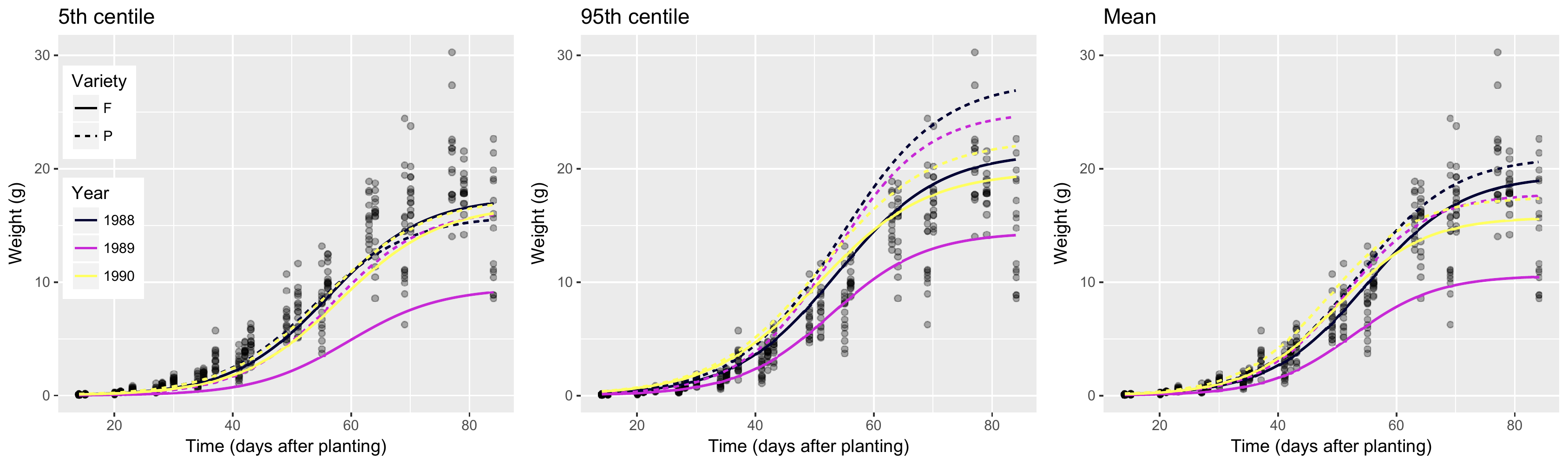}
\caption{Fitted logistic growth curves of soybean plants at the 5th centile (left), 95th centile (center), and at the mean.}
\label{fig:4}
\end{figure}

The three plots in Figure~\ref{fig:4} show the $5$th centile, $95$th centile, and mean predicted growth curves by variety and planting year, while the estimates and standard errors of the fixed effects are reported in Table~\ref{tab:7}. For the sake of brevity, we confine our discussion to the genotypic effect on the asymptote. In 1988, the experimental strain had an advantage over the commercial variety but only at the 95th centile of the leaf weight distribution, with an estimated asymptote difference of $\hat{\beta}_{\tau,4} = 6.31$ g. In the following year, there was a statistically significant interaction between variety and year at the 5th centile, corresponding to an estimated overall effect equal to $\hat{\beta}_{\tau,4} + \hat{\beta}_{\tau,5} = 6.95$ g. The estimated overall effect of variety P on the asymptote at the 95th centile was 10.67 g, thus greater than the effect at the 5th centile and at the mean (7.19 g). Finally, the estimated differences between asymptotes of the growth curves for the experimental and commercial strains in 1990 were negligible: 0.58 g (5th centile), 2.81 g (95th centile) and 0.77 g (mean). In summary, the experimental strain did yield heavier leaves than the F variety, not only in 1989 as estimated by NLME, but also in 1988, and the magnitude of the genotypic effect was dependent on the quantile level.

\begin{table}[h!]
\caption{Estimates of the fixed effects (standard errors) from two logistic quantile mixed-effects models with $\tau \in \{0.05,0.95\}$ and from the normal nonlinear mixed-effects model (NLME) using the Soybean Data. Standard errors for quantile regression estimates are based on 200 bootstrap replications. Bold denotes statistically significant at the $5\%$ level.}
\centering
\begin{tabular}{lrrr}
  \hline
 & $\tau = 0.05$ & $\tau = 0.95$ & NLME \\
  \hline
$\beta_1$ & \textbf{17.49} (1.47) & \textbf{21.43} (2.34) & \textbf{19.43} (0.95) \\
  $\beta_2$ & \textbf{$-$7.99} (1.53) & \textbf{$-$7.02} (2.30) & \textbf{$-$8.84} (1.07) \\
  $\beta_3$ & $-$0.66 (2.06) & $-$1.67 (2.49) & \textbf{$-$3.71} (1.18) \\
  $\beta_4$ & $-$1.64 (2.01) & \textbf{6.31} (1.99) & 1.62 (1.04) \\
  $\beta_5$ & \textbf{8.59} (1.93) & 4.36 (2.41) & \textbf{5.57} (1.17) \\
  $\beta_6$ & 2.22 (2.05) & $-$3.50 (2.01) & 0.15 (1.18) \\
  $\beta_7$ & \textbf{56.16} (1.13) & \textbf{53.71} (2.57) & \textbf{54.81} (0.75) \\
  $\beta_8$ & 3.30 (2.11) & $-$0.86 (2.85) & \textbf{$-$2.24} (0.97) \\
  $\beta_9$ & 1.94 (2.48) & $-$3.14 (2.79) & \textbf{$-$4.97} (0.97) \\
  $\beta_{10}$ & $-$2.50 (1.70) & 0.51 (0.97) & \textbf{$-$1.30} (0.41) \\
  $\beta_{11}$ & \textbf{8.11} (0.32) & \textbf{8.63} (0.79) & \textbf{8.06} (0.15) \\
  $\beta_{12}$ & $-$0.29 (0.51) & $-$0.76 (0.85) & \textbf{$-$0.90} (0.20) \\
  $\beta_{13}$ & 0.40 (0.49) & 0.44 (0.91) & \textbf{$-$0.67} (0.21) \\
   \hline
\end{tabular}
\label{tab:7}
\end{table}

\clearpage

\section{Discussion}
Mixed-effects modeling has a long tradition in statistical applications. There is a vast number of applications of mixed models to the analysis of clustered data in the social, life and physical sciences \citep{pinheiro_bates,demidenko_2013}. Linear quantile mixed models \citep{geraci2007,geraci2014} have too been used in a wide range of research areas, including marine biology \citep{muir_etal_2015,duffy_etal_2015,barneche_2106}, environmental science \citep{fornaroli_etal_2015}, cardiovascular disease \citep{degerud_2014,blankenberg_2016}, physical activity \citep{ng,beets}, and ophthalmology \citep{patel_etal_2015,patel_etal_2016}. The present paper provides a novel and valuable contribution to the modeling of nonlinear quantile functions which broadens the applicability of quantile regression for clustered data.

NLQMMs represent a flexible alternative to nonlinear mixed models for the mean as they allow direct estimation of conditional quantile functions without imposing normal assumptions on the errors. As shown in two real data examples, NLQMMs reveal nonlinear relationships that may be quantitatively and qualitatively different at different quantiles. Also, changes in location, scale, and shape of the response distribution determined by the covariates are naturally brought into light by examining central and tail quantiles \citep{geraci2016}. As compared to nonlinear quantile regression for independent data, our nonlinear estimators are more efficient and they provide additional information about the heterogeneity among clusters.

\section*{Acknowledgements}
This research has been supported by the National Institutes of Health -- National Institute of Child Health and Human Development (Grant Number: 1R03HD084807-01A1).

\appendix
\section*{Appendix}

The estimation algorithm for NLQMM is based on a set of decreasing values of $\omega$. This optimization approach has the appealing advantage of reducing the original problem to an approximated $L_{2}$ problem. The pseudo-code is given below.
\begin{framed}
\begin{center}
\textsc{Smoothing Algorithm with Laplacian Approximation for Nonlinear Quantile Mixed Models}
\begin{itemize}
\item[(1)] Set the maximum number of iterations $T$; the factor $0< \gamma < 1$ for reducing the tuning parameter $\omega$; the tolerance for the change in the log-likelihood; and $t = 0$. Estimate the starting values as follows:
\begin{itemize}
\item[(a)] obtain an estimate for $\bm\beta^{(0)}_{\tau}$ using nonlinear quantile regression \citep{koenker_park}. See, for example, the function \texttt{nlrq} in the \texttt{R} package \texttt{quantreg} \citep{quantreg} which supports self-starting models such as \texttt{SSlogis}. If the nonlinear quantile regression algorithm fails, consider the estimate of the fixed effects from the NLME model in step (1.b) below or, if the latter fails too, a standard nonlinear least squares estimate \citep{bates_watts};
\item[(b)] obtain an estimate for $\bm\xi^{(0)}_{\tau}$ from an NLME model. See, for example, the function \texttt{nlme} in the \texttt{R} package \citep{nlme}. If the NLME algorithm fails, provide an arbitrary value $\bm\xi^{(0)}_{\tau}$;
\item[(c)] obtain an estimate for $\sigma_{\tau}^{(0)}$. For example, this can be estimated as the mean of the absolute residuals from step (1.a) above;
\item[(d)] provide a starting value $\omega^{(0)}$ \citep[see, for example,][p.143]{chen2007};
\item[(e)] using $\bm\beta^{(0)}_{\tau}$, $\bm\xi^{(0)}_{\tau}$, and $\sigma^{(0)}_{\tau}$, solve the penalized least-squares problem \eqref{eq:12} to obtain $\mathbf{u}^{(0)}_{i}$, $i = 1,\dots,M$. See, for example, the \texttt{R} function \texttt{nlm}.
\end{itemize}
\item[(2)] While $t < T$
\begin{itemize}
\item[(a)] Update $\bm\theta_{\tau}^{(t)}$ by minimizing \eqref{eq:13} (or \eqref{eq:14}). See, for example, the \texttt{R} function \texttt{optim}.
\item[(b)] If the change in the log-likelihood is smaller than a given tolerance
\begin{itemize}
    \item[(i)] then return $\bm\theta_{\tau}^{(t + 1)}$;
    \item[(ii)] else set $\bm\theta_{\tau}^{(t + 1)} = \bm\theta_{\tau}^{(t)}$; $\omega^{(t+1)} = \gamma \cdot \omega^{(t)}$; $t = t + 1$; go to step (2.a).
    \end{itemize}
\end{itemize}
\item[(3)] Update $\sigma^{(t)}_{\tau}$ and $\mathbf{u}^{(t)}_{i}$, $i = 1,\dots,M$.
\end{itemize}

\end{center}
\end{framed}

\end{document}